\documentclass[pra,showpacs,preprintnumbers,amsmath,amssymb]{revtex4}
\usepackage{graphicx}
\usepackage{graphics}
\usepackage{amsmath,amsfonts,amssymb}
\usepackage{bm}
\usepackage{dcolumn}
\usepackage{epsfig}

\begin{document}

\title{Effects of nonzero photon momentum in $(\gamma,2e)$ processes\footnote{Submitted to PRA: RAPID COMMUNICATIONS}}

\author{A G Galstyan$^1$\footnote{galstyan@physics.msu.ru}, O Chuluunbaatar$^2$, Yu V Popov$^3$\footnote
             {popov@srd.sinp.msu.ru} and B Piraux$^{4}$}

\affiliation{$^1$ Physics Faculty, Moscow State University, Moscow, Russia\\
                   $^2$ Joint Institute for Nuclear Research, Dubna, Russia\\
                   $^3$ Skobeltsyn Institute of Nuclear Physics, Moscow State University, Moscow, Russia\\
                   $^4$ Institute of Condensed Matter and Nanosciences, Universit\'e catholique de Louvain,\\
                             2, chemin du cyclotron, BOX L7.01.07, B1348 Louvain-la Neuve, Belgium}

\begin{abstract}
We study the effects of nonzero photon momentum on the
triply-differential cross section for ($\gamma$,2e) processes. Due
to the low value of the photon momentum, these effects are weak
and manifest only in special kinematical conditions like the
back-to-back emission of the electrons with equal energy sharing.
Helium and a few light helium-like ions are treated in detail.
Quite unexpectedly, the magnitude of these effects is maximal for
relatively small photon energies. However, although this effect on the TDCS
remains rather small, of the order of a few mbarn $eV^{-1}$
$sr^{-2}$,  it is sufficient to be observed experimentally.
\end{abstract}


\maketitle

\section{INTRODUCTION}

At present, most of the theoretical studies of the interaction of
atoms and molecules with external electric (laser) fields assume
that the vector potential $\vec A$ depends only on time. This, of
course, results from the dipole approximation. This approximation
is usually extremely good. However, in general, $\vec A(\vec r,t)$
depends also on $\vec r$. In the present contribution, we take
into account this dependence by  decomposing  $\vec A(\vec r,t)$
in a basis of plane waves $\exp(\pm\mathrm{i}(\omega t-\vec
k\cdot\vec r))$ where $\omega$ is the photon frequency and $\vec
k$ its momentum. Note that even for one photon transitions, the
vector potential (electric field) depends always on the space
coordinate. However, since $k=\omega/c$, the actual value of $k$
is usually very small so that its effects are expected to manifest
only at rather high frequencies.

To our knowledge, Amusia {\it et al.} were the first to consider
theoretically one-photon double ionization of helium at non
relativistic electron  energies while taking into account the
nonzero photon momentum in their perturbative calculations
\cite{Amusia1}. They found one more process, the so-called
"sea-gull" diagram, which contributes to the amplitude. This
process does occur only if $k\neq 0$. In addition, their
calculations suggest that the amplitude of the effects is
sensitive to the way the helium initial and final state wave
functions behave in the cusp region where the interelectron
distance tends to zero. Unfortunately, at the time their paper was
published, the authors could only treat the electron-electron
interaction in the final state as a perturbation. Recently, Ludlow
{\it et al.} \cite{Colgan} using their time-dependent close
coupling  method, found a peak in the quadrupole energy
distribution of the escaping electrons at 800 eV photon energy.
This peak  that corresponds to equal energy sharing is a signature
of the above process and disappears completely when the photon
momentum is zero. Later on, evidence of this effect has been found
experimentally \cite{exp}.

In this contribution, we use the time independent perturbation
theory to analyze Amusia's process in the case of linearly and
elliptically polarized fields. Contrary to Ludlow {\it et al.},
all multipole interactions are taken into account. In addition,
our helium initial and final state wave functions satisfy the Kato
cusp condition. We calculate the triply-differential cross section
(TDCS) and try to evaluate accurately the relative importance of
the multipole transitions when compared to the dipole ones.

Atomic units $\hbar=e=m_e=1$ are used throughout unless otherwise
specified.

\section{THEORY}

\subsection{Linear polarization}

We have to calculate the following matrix element \cite{Amusia2}
 \begin{equation}\label{1}
M(\vec k)=\int\Psi^{-*}_f(\vec r_1,\vec
r_2)\left[\mathrm{e}^{\mathrm{i}\vec k\cdot\vec
r_1}(\vec\epsilon\cdot{\hat{\vec p}}_1)+\mathrm{e}^{\mathrm{i}\vec
k\cdot\vec r_2}(\vec\epsilon\cdot{\hat{\vec
p}}_2)\right]\Psi_i(\vec r_1,\vec r_2)d^3r_1d^3r_2.
 \end{equation}
Here ${\hat{\vec p}}_j\equiv-\mathrm{i}\vec \nabla_j$, $j=1,2$.
The functions $\Psi$ are wave functions solution of the field free
helium Hamiltonian. For clarity, we define the coordinates of all
vectors as follows: $\vec\epsilon=\{0,0,1\}$ is the unit vector
along the polarization axis that coincides with the $z$-axis; the
outgoing electron momenta are $\vec
p_1=p_1\{\sin\theta_1\cos\varphi_1,\sin\theta_1\sin\varphi_1,\cos\theta_1\}$
and $\vec
p_2=p_2\{\sin\theta_2\cos\varphi_2,\sin\theta_2\sin\varphi_2,\cos\theta_2\}$
and the photon momentum $\vec k=
{\omega}/{c}\{\cos\phi;\sin\phi;0\}$. The energy conservation
writes $\omega+\varepsilon_0^{He}= p_1^2/2+p_2^2/2$ where
$\varepsilon_0^{He}$ is the helium ground state energy. The TDCS
is given by ($\alpha=c^{-1}=1/137$):
\begin{equation}\label{2}
\frac{d^3\sigma}{d\Omega_1d\Omega_2dE_1}=2\frac{\alpha
p_1p_2}{(2\pi)^4\omega}\left[\frac{1}{2\pi}\int\limits_0^{2\pi}|M(\vec
k)|^2d\phi\right].
\end{equation}
In most of the experimental geometries and kinematical conditions,
$k$ is rather small compared to the electron momenta. As a result,
it is straightforward to show that in very good approximation,
 \begin{equation}\label{3}
M(\vec k)=M(0)+\left[(\vec k\cdot\vec p_1)g_1+(\vec k\cdot\vec
p_2)g_2\right]+O(k^2).
 \end{equation}
Eq. (\ref{3}) shows that an effect of the nonzero photon momentum
can be seen in the angular domain where $M(0)\sim 0$. It is well
known \cite{Briggs} that, for equal energy sharing, {\it i.e.}
when $p_1=p_2$, the matrix element $M(0)=0$ if $\vec p_1=-\vec
p_2$ (back-to-back emission, $\theta_2=\pi-\theta_1$,
$\varphi_2=\pi+\varphi_1$). Let us consider the coplanar case,
when both momenta are disposed in the plane $(x,z)$: $\vec
p_1=p_1\{\sin\theta,0,\cos\theta\}$ and $\vec p_2=p_1
\{-\sin\theta,0,-\cos\theta\}$. In this geometry, it is now
possible to separate the $\phi$ and $\theta$ dependence and write:
 \begin{equation}\label{4}
M(\vec k)\approx k\cos\phi G(\theta),
 \end{equation}
where the expression of $G(\theta)$ is, for the time being,
unspecified. Averaging on the variable $\phi$ in (\ref{2})  is
trivial: $\overline{|M(\vec k)|^2}=1/2 |M(\vec k)_{\phi=0}|^2$. In
the general case, $\overline{|M(\vec k)|^2}\approx |M(0)|^2+ 1/2
|M(\vec k)_{\phi=0}-M(0)|^2$.\\

In order to estimate the magnitude of the TDCS, we use the  same initial and
final wave functions considered in the model described earlier in
\cite{Chuka10}. The properly normalized  and correlated initial state wave
function is given by:
\begin{equation}\label {5}
\Psi_i(\vec r_1,\vec r_2)=
\sum_{j}D_j(\mathrm{e}^{-a_jr_1-b_jr_2}+\mathrm{e}^{-a_jr_2-b_jr_1})\mathrm{e}^{-\gamma_jr_{12}},
\end{equation}
giving the helium ground energy $\varepsilon_0^{He}=-2.90372$ a.u.. The
final double continuum wave function is given by the well known 3C
function:
 \begin{equation}\label {6}
\Psi^{(-)}_f(\vec r_1,\vec r_2)=\mathrm{e}^{\mathrm{i}\vec p_{12}\cdot\vec
r_{12}}\phi^{-*}_1\phi^{-*}_2\phi^{-*}_{12}.
\end{equation}
Here
$$
\phi_j^{-*}(\vec p_j, \vec r)=R(\xi_j)e^{-\mathrm{i}\vec p_j\vec
r}{_1}F_1[-\mathrm{i}\xi_j,1;\mathrm{i}(p_jr+\vec p_j\cdot\vec r)],
$$
with
$$
\vec p_{12}=\frac12(\vec p_1-\vec p_2); \quad
\xi_{12}=\frac{1}{2p_{12}}; \quad \xi_j=-\frac{2}{p_j}\
(j=1,2);\quad R(\xi)=e^{-\pi\xi/2}\ \Gamma(1+\mathrm{i}\xi).
$$
In these conditions, we are left with the evaluation of the following 3-dimensional integral:
\begin{eqnarray}
M(\vec k)&=&\mathrm{i}\sum_{j}D_j\int\frac{d^3p}{(2\pi)^3}
        \{\gamma_jK_{12}(\vec p_{12};\vec p_{12}-\vec p;\vec \epsilon;\gamma_j)\times\nonumber\\
 &&\left[\frac{\partial I_1(\vec p_1;\vec k+\vec p;a_j)}{\partial a_j}
        \frac{\partial I_2(\vec p_2;-\vec p;b_j)}{\partial b_j}
       - \frac{\partial I_2(\vec p_2;\vec k-\vec p;a_j)}{\partial a_j}
        \frac{\partial I_1(\vec p_1;\vec p;b_j)}{\partial b_j}+(a_j\rightleftarrows b_j)\right]+
        \frac{\partial I_{12}(\vec p_{12};\vec p_{12}-\vec p;\gamma_j)}{\partial\gamma_j}\times\nonumber\\
& & \left[a_jK_1(\vec p_1;\vec k+\vec
p;\vec\epsilon;a_j)\frac{\partial I_2(\vec p_2;-\vec
p;b_j)}{\partial b_j}+
        a_jK_2(\vec p_2;\vec k-\vec p;\vec\epsilon;a_j)\frac{\partial I_1(\vec p_1;\vec p;b_j)}{\partial b_j}+
        (a_j\rightleftarrows b_j)\right]\}.
\end{eqnarray}
In (7),
\begin{equation}
I_x(\vec p_x,\vec p,\lambda)=\int\frac{d^3r}{r}\mathrm{e}^{\mathrm{i}\vec p\cdot\vec
r}\phi_x^{-*}(\vec p_x,\vec r)\mathrm{e}^{-\lambda r} = 4\pi\
R(\xi_x)\frac{[(\lambda-\mathrm{i}p_x)^2+p^2]^{\mathrm{i}\xi_x}}{[(\vec p-\vec
p_x)^2+\lambda^2]^{(1+\mathrm{i}\xi_x)}},
\end{equation}
and
\begin{eqnarray}
K_x(\vec p_x,\vec p,\vec e,\lambda)&=&\int\frac{d^3r}{r}\mathrm{e}^{\mathrm{i}\vec
p\cdot\vec r}\phi_x^{-*}(\vec p_x,\vec r)(\vec e\cdot\vec r)\mathrm{e}^{-\lambda r}\nonumber\\
&=&-2\mathrm{i}I_x(\vec p_x,\vec p,\lambda)\left[\mathrm{i}\xi_x\frac{\vec e\cdot\vec
p}{(\lambda-\mathrm{i}p_x)^2+p^2}-(1+\mathrm{i}\xi_x)\frac{\vec e\cdot(\vec p-\vec
p_x)}{(\vec p-\vec p_x)^2+\lambda^2}\right].
\end{eqnarray}
The first term in the figure brackets in the rhs of Eq. 7 just
corresponds to the "sea-gull" graph discussed in \cite{Amusia1}
and disappears if $k=0$.

\subsection{Elliptic polarization}

In the case of elliptically polarized photons, we have $\vec
\epsilon=\{\mathrm{i}\sin\beta,0,\cos\beta\}$ with $-\pi/2\leq\beta\leq\pi/2$, and
$\vec k=\alpha\omega\{0,1,0\}$. The case $\beta=0$ corresponds to the
linear polarization.

\section{RESULTS AND DISCUSSION}

In Fig.1, we consider the case $k=0$ and compare our results for the
absolute TDCS as a function of $\theta_2$ with the data of two experiments, by Schwartzkopf
and Schmidt \cite{exp1} and Brauning {\it et al.} \cite{exp2}. In both cases,
the photon energy is equal to 99 eV, the two electrons share the
same energy, $E_1=E_2=10$ eV and $\vec p_1$ is along the polarization axis.
\begin{figure}[h]
\begin{center}
\end{center}
\includegraphics[width= 10cm,height = 8cm]{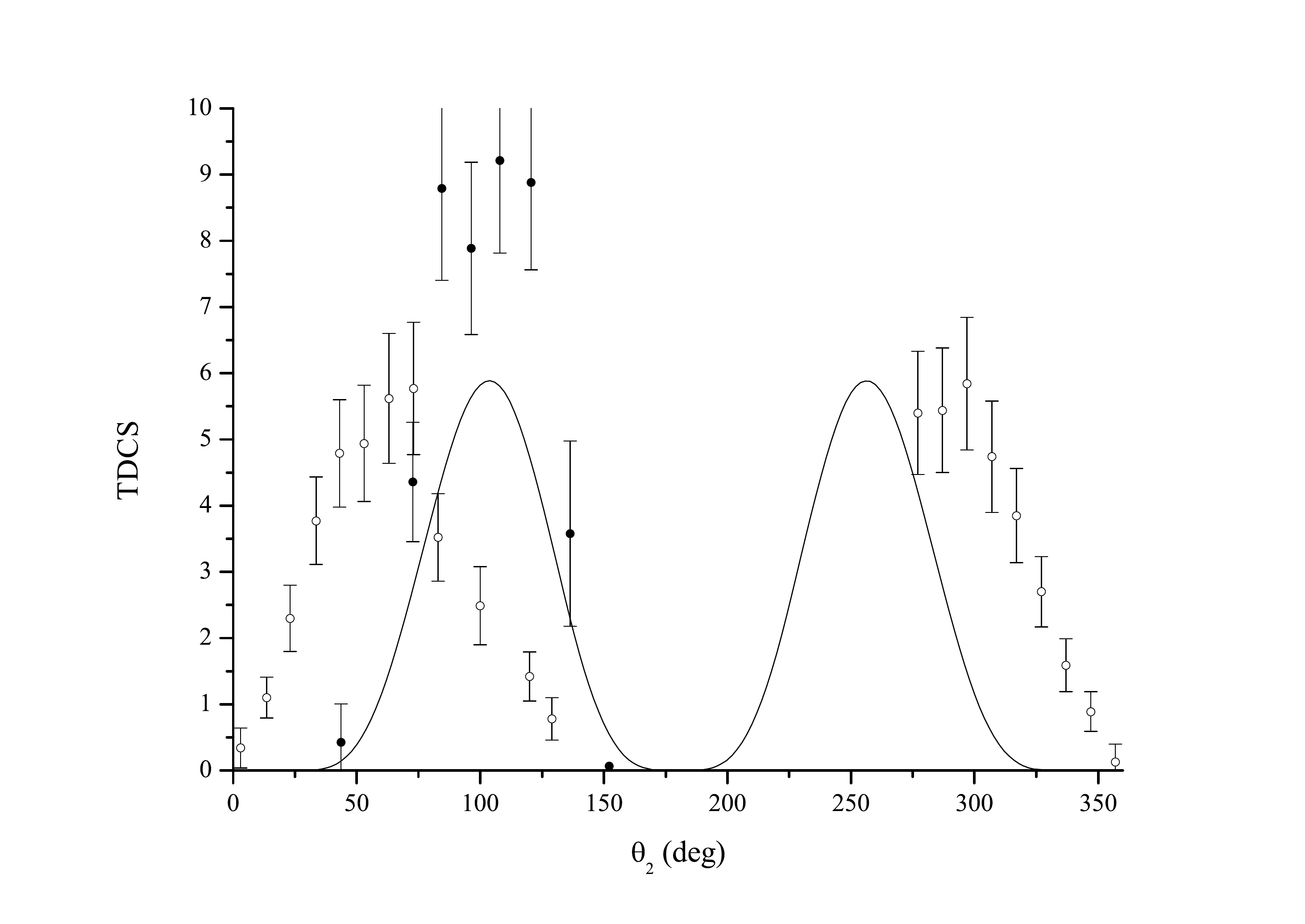}
\caption{Absolute TDCS in b eV$^{-1}$ sr$^{-2}$ (given by (\ref{2})) as a function
of $\theta_2$, the angle between $\vec p_2$ and the polarization axis. The photon
momentum $k=0$, $\omega=99$\ eV, $\vec p_1$ is directed along the polarization
axis and $E_1=E_2=10$\ eV. Our results (full line) are compared to the
data of two experiments: open circles, Schwartzkopf and Schmidt \cite{exp1} and black dots,
Brauning {\it et al.} \cite{exp2}.}
\end{figure}
The qualitative agreement of our results with the experimental
data of Brauning {\it et al.} and the fact that it is also the
case with other theoretical approaches that reproduce the correct
peak position seem to suggest that there is a problem with the
experimental results of Schwartzkopf and Schmidt. On the other
hand, it is legitimate to expect that the present model gives at
least reliable qualitative results.
\begin{figure}[h]
\begin{center}
\includegraphics[width= 12cm,height = 8cm]{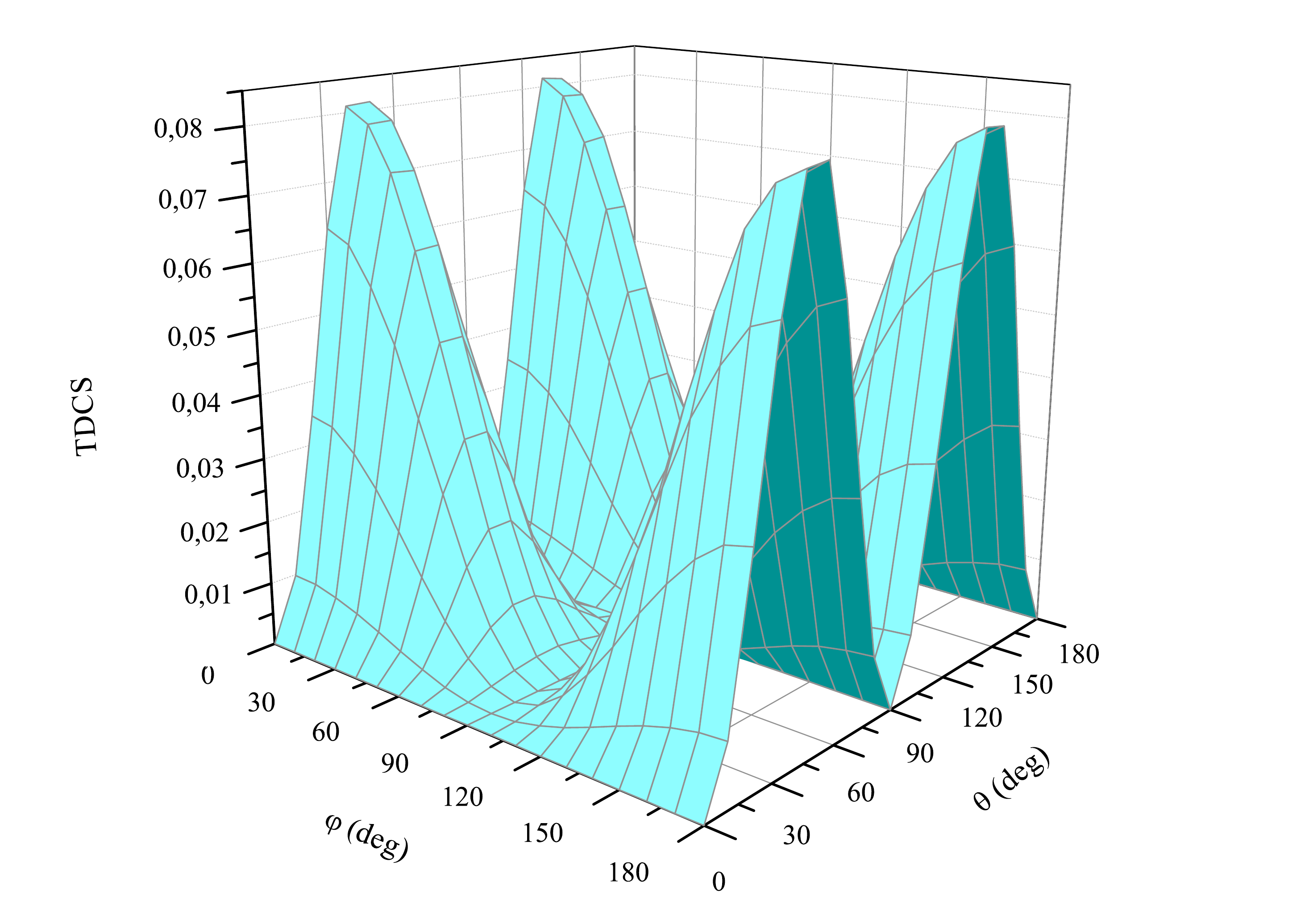}
\end{center}
\caption{(Color online) Absolute averaged TDCS in mb eV$^{-1}$
sr$^{-2}$ in the case of a back-to-back electron emission. $\vec
p_1=-\vec
p_2=p\{\sin\theta\cos\varphi,\sin\theta\sin\varphi,\cos\theta\}$,
\ $\omega=799$\ eV, $E_1=E_2=360$\ eV and $k=\alpha\omega$. }
\end{figure}
In Fig.2, we consider the case of a back-to-back electron emission
($\vec p_1=-\vec p_2$). The averaged TDCS is shown as a function
of the angles $\theta$ and $\varphi$  of the escaping
electrons. The kinetic energy of each electron is $E=360$ eV. Note
that for a back-to-back emission, the electron energy distribution
presents a local maximum at equal energy sharing \cite{Colgan} (see
also Fig. 6 taking into account that $M(0)=0$). We clearly see a
4-peak angle distribution. This contrasts with the zero photon
momentum case where this distribution is uniformly zero. Of
course, the effect is very small, the magnitude of it being of the
order of a fraction of millibarn.
\begin{figure}[h]
\begin{center}
\includegraphics[width= 9cm,height = 7.5cm]{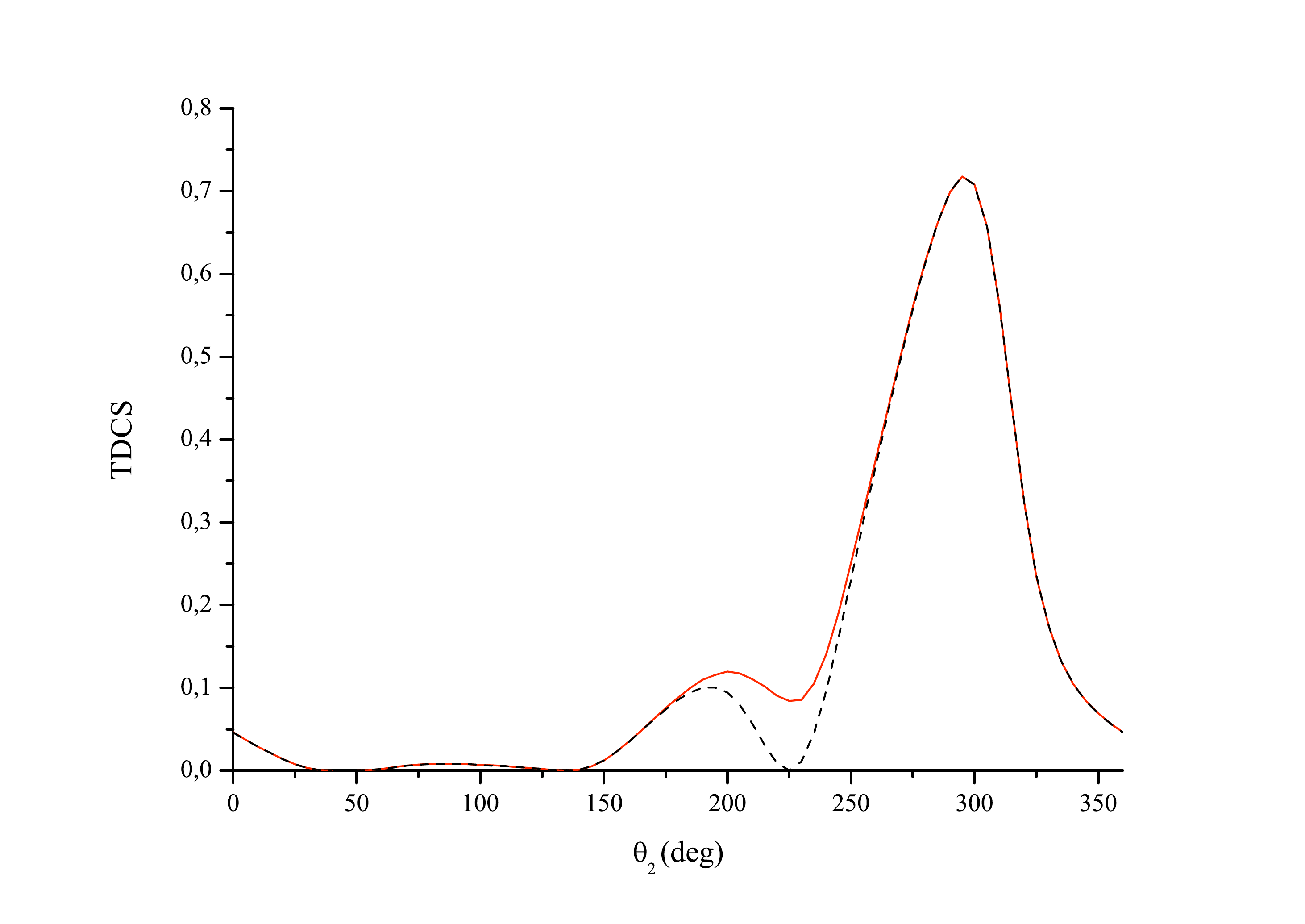}
\end{center}
\caption{(Color online) Absolute TDCS in mb eV$^{-1}$ sr$^{-2}$
(given by (2) and therefore averaged over the variable $\phi$) as
a function of $\theta_2$ for $\omega=799$\ eV and $E_1=E_2=360$\
eV. It is assumed that the angle $\theta_1$ between the momentum
$\vec p_1$ and the polarization axis is fixed to $45^{\circ}$. In
addition, the direction of the vector $\vec p_2$ rotates with
$\theta_2$ in the plane formed by $\vec p_1$ and the polarization
axis. Two different values of the photon momentum $k$ are
considered: $k=0$ (dotted line) and $k=\alpha\omega$ (full line).}
\end{figure}
This effect is also observable in Fig. 3 for the same case as in
Fig. 2. Here, however,  the angle between the electron momentum $
\vec p_1$ and the polarization vector is fixed at 45$^{\circ}$.
These two vectors form a plane in which $\vec p_2$ rotates. The
TDCS is shown as a function of $\theta_2$ for two values, namely 0
and $\alpha\omega$ of the photon momentum $k$. The peak at
$\theta_2=215^{\circ}$ for $k=\alpha\omega$ is about twice higher
than the value obtained for $k=0$ but again, the effect is rather
small.\\

\begin{figure}[h]
\begin{center}
\includegraphics[width= 16cm,height = 10cm]{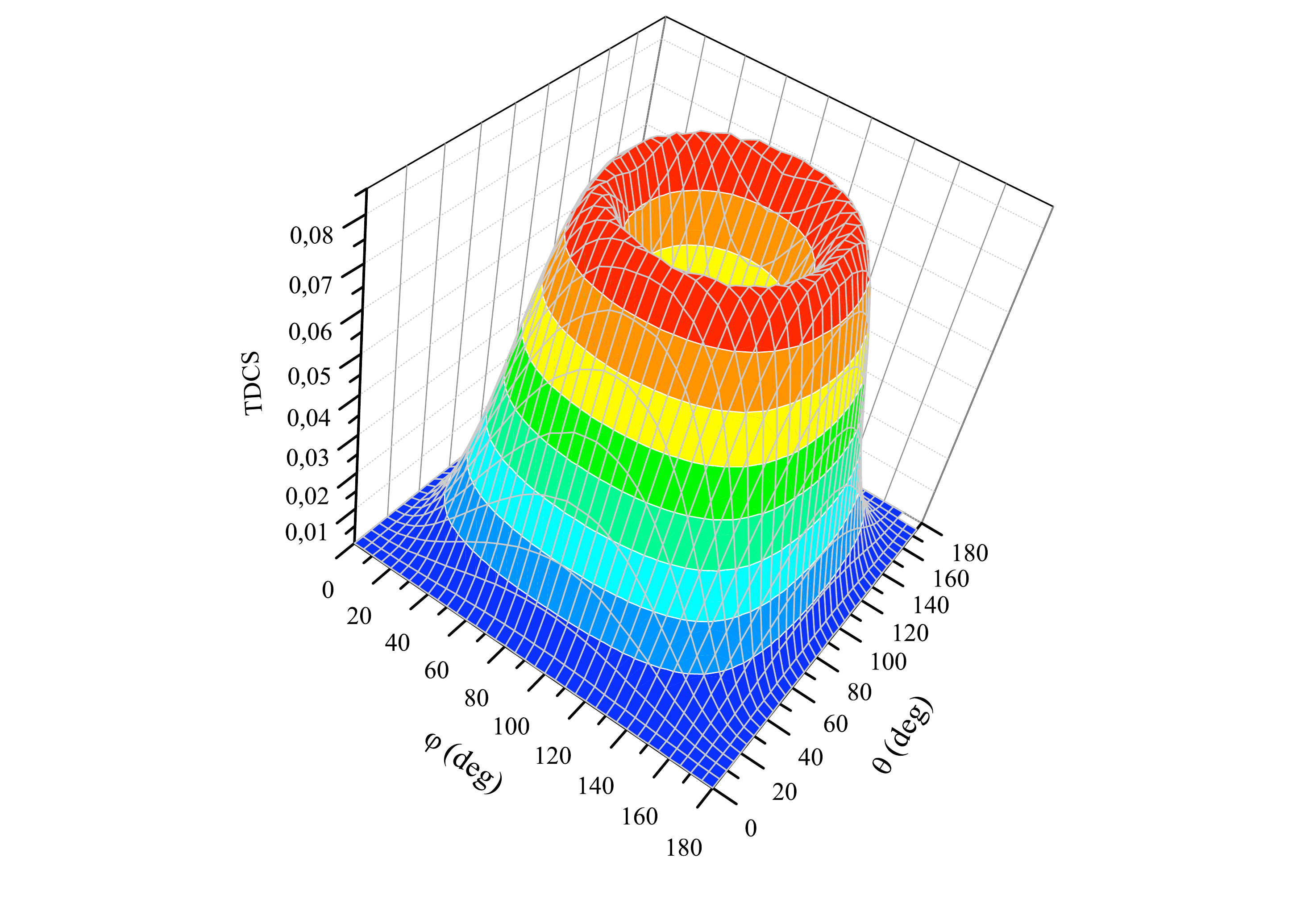}
\end{center}
\caption{(Color online) Absolute TDCS in mb eV$^{-1}$ sr$^{-2}$ in
the case of a back-to-back electron emission. Angles $\theta$ and
$\varphi$ are the same like in Fig. 2, \ $\omega=799$\ eV,
$E_1=E_2=360$\ eV and $k=\alpha\omega$. In this case, the field is
circularly polarized, $\beta=45^{\circ}$.}
\end{figure}
\begin{figure}[h]
\begin{center}
\end{center}
\includegraphics[width= 10cm,height = 6cm]{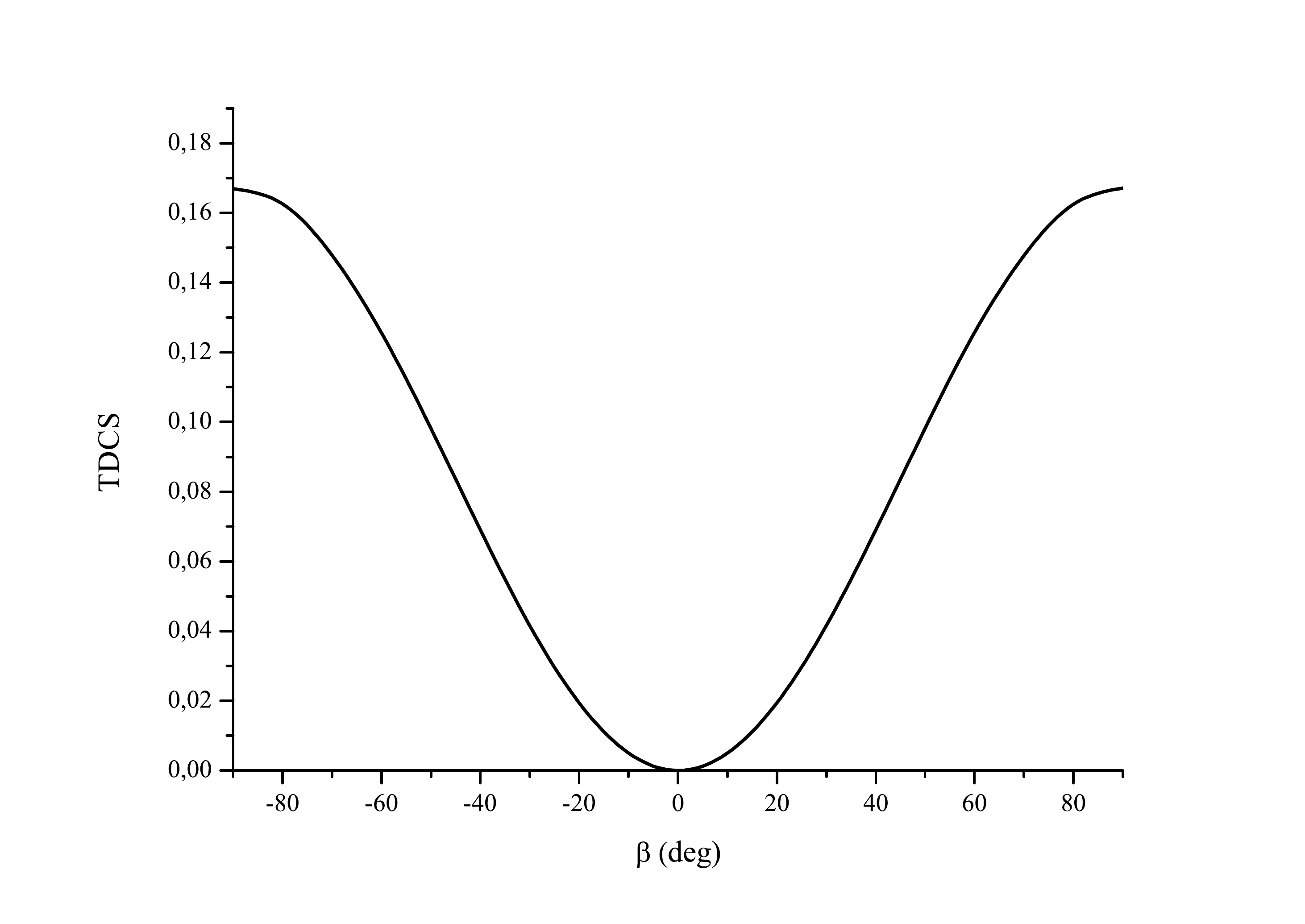}
\caption{Absolute TDCS in mb eV$^{-1}$ sr$^{-2}$ for
$\theta=90^{\circ},\ \varphi=135^{\circ}$ as a function of $\beta$
that  determines the polarization of the field. As in Fig. 2, the
photon energy is equal to 799 eV, $k=\alpha\omega$ and it is
assumed that both electrons are emitted back-to-back with
$E_1=E_2=360$ eV.}
\end{figure}
\begin{figure}[h]
\begin{center}
\includegraphics[width= 10cm,height = 8cm]{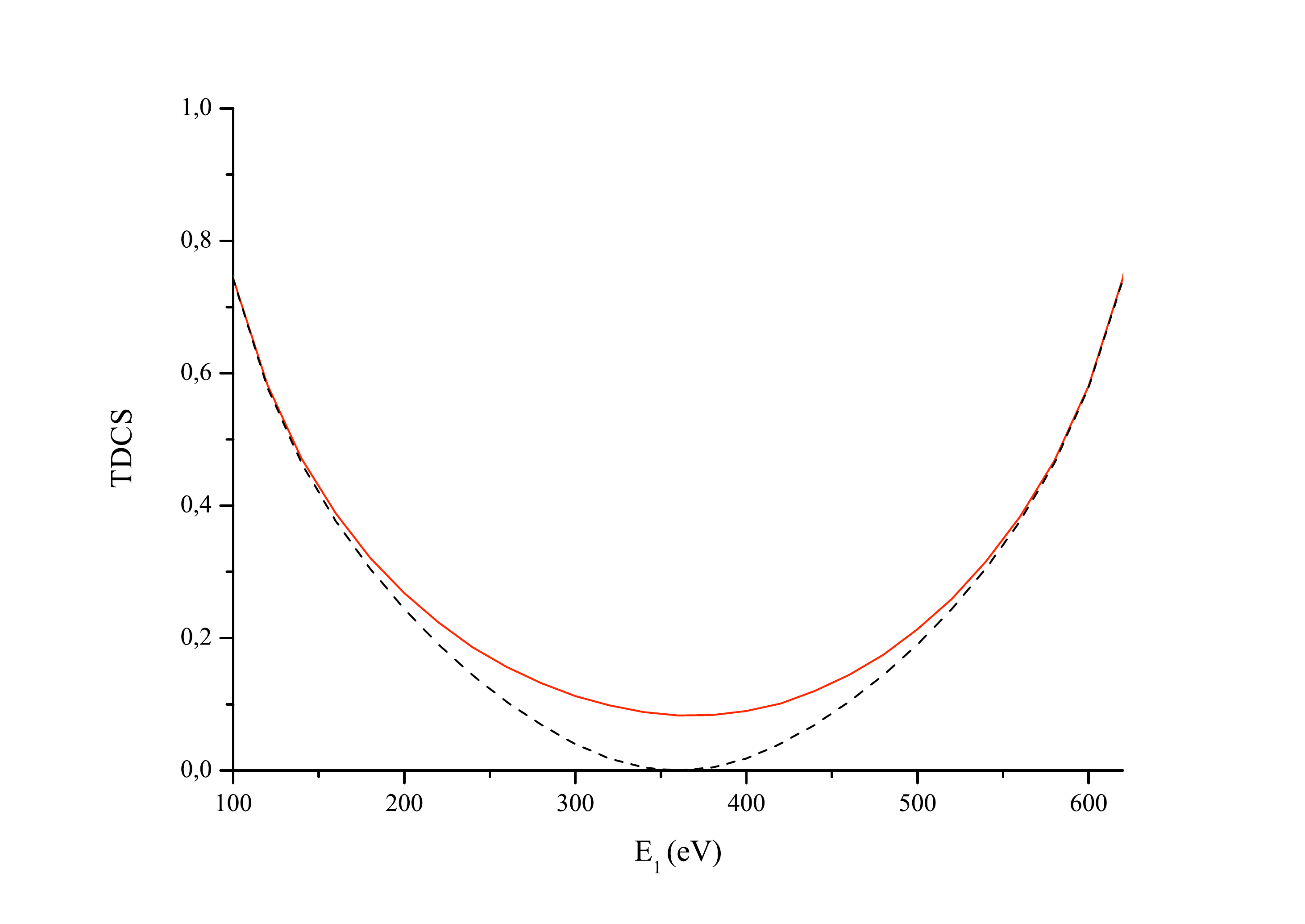}
\end{center}
\caption{(Color online) Electron energy distribution of the TDCS
in mb eV$^{-1}$ sr$^{-2}$. Electrons move in opposite directions
with $\theta=45^{\circ}$, $\varphi=0$ and $\omega=799$\ eV. Solid
line, $k=\alpha\omega$, dashed line, $k=0$.}
\end{figure}
\begin{figure}[h]
\begin{center}
\includegraphics[width= 14cm,height = 12cm]{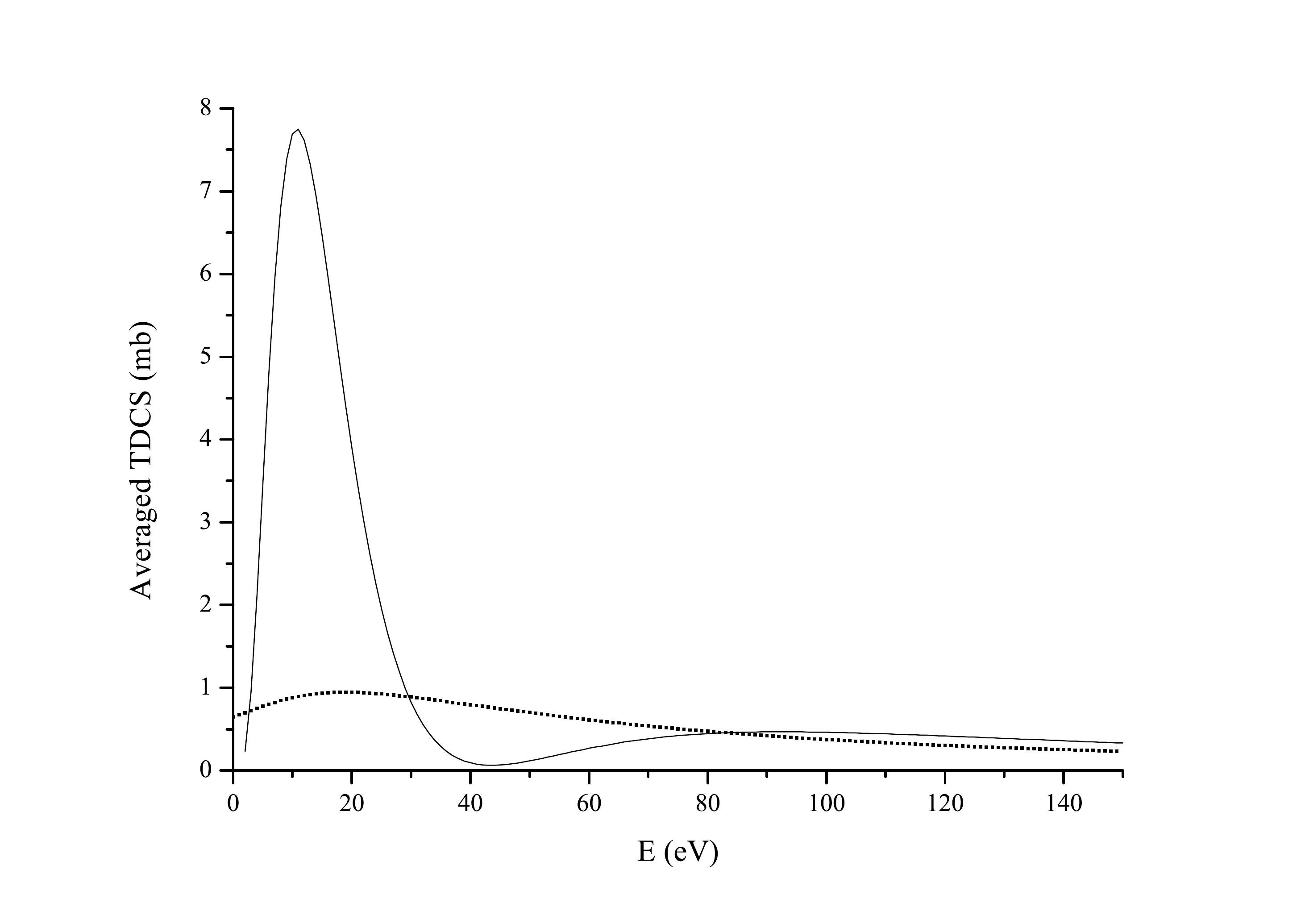}
\end{center}
\caption{Dependence of the maximum of the TDCS (in mb eV$^{-1}$
sr$^{-2}$) on the energy $E=E_1=E_2$ of the outgoing electrons
moving in opposite directions with $\theta=45^\circ,\ \varphi=0$
and $k=\alpha\omega$. The solid line corresponds to a 3C final
state wave function and the dotted line to the uncorrelated 2C
final state wave function.}
\end{figure}
So far, we have considered the case of a linear polarization. Let
us now assume the field circularly polarized.  In Fig. 4, we
consider the same case as in Fig. 2 except that $\beta$ which
determines the type of field polarization is now equal to
$45^{\circ}$, what corresponds to the circular polarization. The
TDCS as a function of $\theta$ and $\varphi$ exhibits a volcano
type of structure whereas for zero photon momentum, it is
uniformly zero. Note that for $\beta=0$ (linear polarization),
this distribution reduces to the one given in Fig. 2. It is
interesting to analyze for what value of $\beta$, the effects due
to nonzero photon momentum is the strongest. In Fig. 5, we show
the TDCS for $\theta=90^{\circ}$ and $\varphi=135^{\circ}$ as a
function of $\beta$ for the same case as before, namely,
$k=\alpha\omega$,  a photon energy of 799 eV and $E_1=E_2=360$ eV.
We clearly see that the TDCS reaches its highest value for
$|\beta|\gtrsim 75^{\circ}$, {\it i.e.} for a highly elliptical
polarization. It is interesting to note that if we were able to
create a beam of such highly elliptically polarized photons, we
would practically double the TDCS in comparison to the linear
polarization. Note that for $\beta=0$ (linear polarization), the
TDCS is equal to zero. This results from the fact that
$\theta=90^{\circ}$ and $\varphi=135^{\circ}$ (see Fig. 2).

Let us now study the electron energy distribution in the case the
field polarization is linear. As before, we assume that both
electrons are emitted back-to-back and consider the same case as
in Fig. 2 namely a photon energy of 799 eV and a photon momentum
given by $k=\alpha\omega$. We also set $\theta=45^{\circ}$ and
$\varphi=0$. This corresponds to a maximum of the TDCS in Fig. 2.
The results for the energy distribution are shown in Fig. 6 where
they are compared with those obtained with $k=0$. We clearly
observe an effect resulting from the nonzero value of the photon
momentum for $E_1=E_2=360$ eV. Note however that at $E_1=E_2=E$,
the TDCS does not exhibit a local maximum. This is because, there
is no integration on the solid angles $\Omega_1$ and $\Omega_2$.

In Fig. 7, we study how the maximal value of the TDCS that occurs
at $\theta=45^{\circ}$ and $\varphi=0$ (see Fig. 2) varies with
the energy $E$ of each electron or, in other words, with the
photon energy. As before, it is assumed that both electrons are
emitted back-to-back. Quite unexpectedly, we clearly see that the
amplitude of the maximum of the TDCS occurs at the rather low
energy $E\approx 10$ eV that corresponds to  a photon energy of
about 100 eV. In addition, the results presented in Fig. 7
demonstrate clearly that the amplitude of the effects due to the
nonzero photon momentum depends strongly on the way the final
state electron correlation is treated. In the case of the 3C
function, the amplitude of the maximum of the TDCS near $E=10$ eV
is almost eight times the value obtained by using an uncorrelated
2C function to describe the final state wave function. Note that
for values of $E>80$ eV, both functions lead to very similar
results as expected. This dependence of the TDSC on the final
state electron correlations confirms what was already suggested in
Amusia's work \cite{Amusia1}.

\begin{table}[h]
\caption{Energy $E=E_1=E_2$ of the escaping electrons leading to a
maximum value of the TDCS assuming a back-to-back electron
emission with $\theta=45^{\circ}$ and $\varphi=0^{\circ}$. Various
nucleus charges $Z$ of helium-like ions are considered.}
\vspace{0.5cm}
\begin{tabular}{p{4cm}p{4cm}p{4cm}}\hline\\
$Z$ & $E$ (eV) & maximum of TDCS (mb)\\ \\ \hline\\
2 & 11 & 7,75 \\
3 & 21 & 1,45 \\
4 & 33 & 0,47 \\
5 & 45 & 0,19 \\
6 & 75 & 0,10 \\ \\ \hline
\end{tabular}
\end{table}

Within the present context, it is legitimate to ask whether helium
is the best target to observe effects due to nonzero photon
momentum. In order to answer to that question, let us consider
helium-like ions and examine the influence of the nucleus charge
$Z$ on the relative amplitude of the effects of nonzero photon
momentum on the TDCS. We assume the external field linearly
polarized and consider a back-to-back electron emission with
$\theta=45^{\circ}$, and $\varphi=0^{\circ}$ since this
configuration leads  to a maximum of the TDSC for $k\ne 0$ as
shown in Fig. 2. The ground states wave functions for a few light
helium-like ions are generated following a procedure described in
\cite{Chuka06}. Our results are presented in Table 1. For each
ion, we calculate the energy of the escaping electrons that lead
to a maximum of the TDCS in the above configuration. We clearly
see that for increasing values of the nucleus charge $Z$, the
absolute value of the maximum of the TDCS decreases rapidly.
Taking into account that the effects due to nonzero photon
momentum are very small, it follows that from the experimental
point of view, only  helium and may be lithium are suitable
targets.

\section{CONCLUSIONS}

In this contribution, we considered $(\gamma,2e)$ processes and
studied in detail the effects resulting from  nonzero value of
the photon momentum on the triply-differential cross section. Due
to the small value of the photon momentum, these effects are quite
small, of the order of a few mb, and are only observable in
particular kinematical conditions. It is the case when both
electrons are emitted back-to-back with equal energy sharing. In fact, for
this configuration, the wave function has a node when the photon
momentum is identically zero (dipole approximation). In these
conditions, the absolute value of the effect depends on the energy
and the polarization of the photon. The effect is the strongest
when the polarization is linear or strongly elliptical.
Furthermore and quite unexpectedly, we have shown that the effect
is maximal for relatively low photon energies. We have also shown
that the  final state electron correlations play an important
role. Neglecting electron correlations in the final state leads to
a quite severe underestimation of the amplitude of the effect.
Finally, we examined the case of several helium-like ions and
showed that in the same kinematical conditions, helium or may be
lithium are more suitable targets for observing experimentally these
effects due to nonzero photon momentum.

\section*{Acknowledgements}

We are grateful to Th. Weber for stimulating discussions. Yu.V.P.
thanks the Universit\'e catholique de Louvain for hospitality and
financial support. A.G., O.Ch. and Yu.V.P. acknowledge the Russian
Foundation for Basic Research (grant no. 11-01-00523) for
financial support. O. Ch. thanks a financial support from the
theme 09-6-1060-2005/2013 ``Mathematical support of experimental
and theoretical studies conducted by JINR''. We also express our
gratitude to Faculty of Computer Mathematics and Cybernetics of
Moscow State University for providing access to the supercomputer
Blue Gene/P.


\end{document}